\documentclass[12pt]{iopart}

\usepackage{iopams}

\usepackage{graphicx}
\usepackage{color}
\usepackage{braket}
\newcommand{\op}[1]{\hat{ #1}}

\usepackage{url}
\usepackage[dvipdfmx]{hyperref}
\hypersetup{
setpagesize=false,
bookmarksnumbered=true,%
bookmarksopen=true,%
colorlinks=true,%
linkcolor=blue,
citecolor=blue,
urlcolor=blue
}

\usepackage{cite}

\usepackage{multirow}

\begin{document}

\title{Implementation of quantum state tomography for time-bin qudits}

\author{Takuya Ikuta and Hiroki Takesue}

\address{NTT Basic Research Laboratories, NTT Corporation, 3-1 Morinosato Wakamiya, Atsugi, Kanagawa 243-0198, Japan}
\ead{ikuta.takuya@lab.ntt.co.jp}
\vspace{10pt}
\begin{indented}
\item[]\today
\end{indented}

\begin{abstract}
Quantum state tomography (QST) is an essential tool for characterizing an unknown quantum state.
Recently, QST has been performed for entangled qudits based on orbital angular momentum,
time-energy uncertainty, and frequency bins.
Here, we propose a QST for time-bin qudits,
with which the number of measurement settings scales linearly with dimension $d$.
Using the proposed scheme,
we performed QST for a four-dimensional time-bin maximally entangled state with 16 measurement settings.
We successfully reconstructed the density matrix of the entangled qudits,
with which the average fidelity of the state was calculated to be 0.950.
\end{abstract}

\pacs{03.65.Wj, 03.65.Ud, 42.50.Dv}
%
\noindent{\it Keywords}: quantum state tomography, time bin, entanglement

\submitto{\NJP}
%
%
%

\section{Introduction}

In quantum information science,
many figures of merit such as fidelity and von Neumann entropy \cite{Nielsen2010} are utilized to characterize a quantum state.
Quantum state tomography (QST) \cite{James2001},
by which a quantum density operator of an unknown quantum state is identified,
is the most comprehensive method for deriving them.
Recently, QST for photonic high-dimensional quantum states (qudits) \cite{Thew2002} has been intensively investigated for
entanglements based on orbital angular momentum \cite{Agnew2011},
frequency bins \cite{Bernhard2013},
and time-energy uncertainty \cite{Richart2014}.
Observation of high-dimensional multipartite entanglement has also been reported \cite{Malik2016}.
For time-bin qudits, which are promising candidates for transmission over an optical fiber,
QST based on the conversion between time-bin states and polarization states has been performed \cite{Nowierski2015}.
QST generally requires $(d^2 - 1)$ different measurements for a state in $d$ dimensional Hilbert spaces
because a general mixed state is characterized by $(d^2 - 1)$ real numbers.
Thus, it is important to reduce the number of measurement settings for high-dimensional QST.
For time-bin qubits,
QST has been performed with a single delay Mach-Zehnder interferometer (MZI) \cite{Takesue2009},
which simultaneously constructed measurements projecting on two time-bin basis states and a superposition state of the time-bin basis.
In this paper, we propose an efficient scheme to implement QST for time-bin qudits utilizing cascaded delay MZIs \cite{Ikuta2016a,Richart2012}.
Thanks to the simultaneous construction of the different measurements,
the number of measurement settings scales linearly with dimension $d$.

\section{\label{sec:QSTDetail}Measurements with cascaded MZIs}
\subsection{\label{sec:BasicQST}Basic concept}

First, we give a general description of QST.
A $d$-dimensional density operator $\op{\rho}$ can be expressed as $\op{\rho} = \sum_{i=0}^{d^2-1} g_i \op{G}_i$,
where $\op{G}_i$ is the generalized Gell-Mann matrix defined in \cite{Thew2002}
and $g_i$ is a real number.
$g_0$ is usually fixed to $1/d$ to be $\mathrm{Tr}\left( \op{\rho} \right) = 1$,
because $\op{G}_i$ is traceless for $i \geq 1$ and $\op{G}_0$ is the identity operator $\op{I}_d$.
When we repeat a measurement represented by a projector $\op{P}_j$ for $N$ photons,
the expected values of the photon counts $n_j^E$ is given by
\begin{equation}
n_j^E = N \mathrm{Tr} \left( \op{P}_j \op{\rho} \right) = N \sum_{i = 0}^{d^2 - 1} A_{ij} g_i	,	\label{eq:ConceptOfQST}
\end{equation}
where $A_{ij} = \mathrm{Tr} \left( \op{P}_j \op{G}_i \right)$.
We can estimate $N$ and $g_i$ by multiplying the inverse matrix of $A_{ij}$ from the left of \eref{eq:ConceptOfQST}.
Thus, the problem remaining to complete QST is how to prepare a set of measurements
that correspond to $\op{P}_j$ for constructing $A_{ij}$ with rank $d^2$.

\begin{figure}[htbp]
\centering
\includegraphics[width=\linewidth]{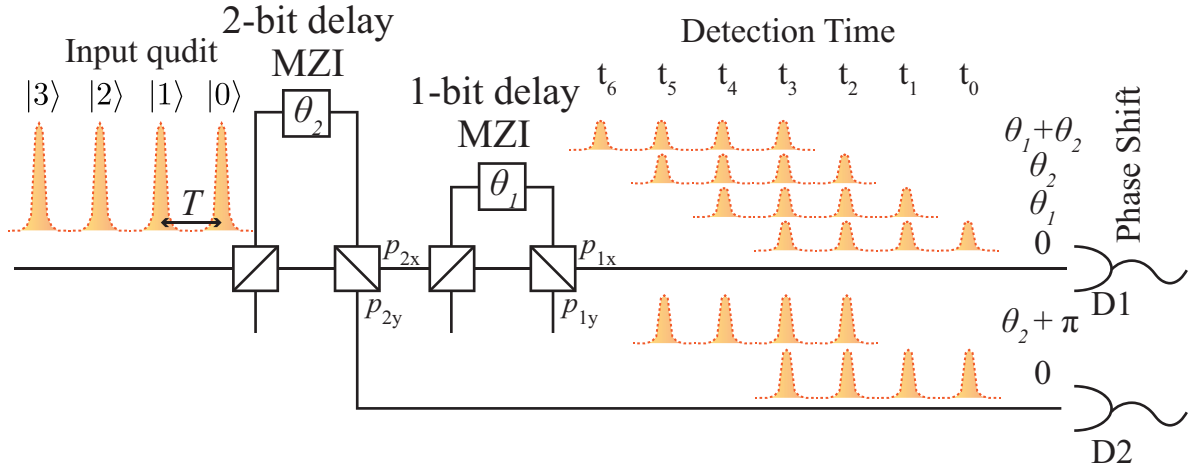}
\caption{Concept of QST utilizing cascaded MZIs.}
\label{fig:CMZI}
\end{figure}

To prepare such a set of measurements for time-bin qudits,
we use cascaded MZIs.
\Fref{fig:CMZI} shows the concept of the measurements with the cascaded MZIs for a four-dimensional time-bin state.
The 2-bit delay MZI has time delay $2T$ and phase difference $\theta_2$,
where $T$ denotes the temporal interval of time slots constituting the time-bin basis
and $\theta_2$ is the phase difference between the short and the long arms of the 2-bit delay MZI.
The 1-bit delay MZI has time delay $T$ and phase difference $\theta_1$.
The output ports of the 2-bit delay MZI, $p_{2x}$ and $p_{2y}$, are connected to the input port of the 1-bit delay MZI and photon detector D2, respectively.
The output port of the 1-bit delay MZI, $p_{1x}$, is connected to photon detector D1,
and the other output port, $p_{1y}$, is terminated.
When the time-bin qudit is launched into the cascaded MZIs,
D1 can detect a photon in a superposition of four different input states.
On the other hand,
D2 cannot,
but it can detect a photon projected on the time-bin basis, which D1 cannot.
Therefore, the information obtained from D1 and D2 are intrinsically different.
We utilize the number of photons detected by D1 and D2 at different detection times as $n_j^E$ in \eref{eq:ConceptOfQST}.

In what follows, we describe the measurements by the cascaded MZIs in more detail.
The basis for the four-dimensional time-bin state is given by state $\ket{k} (k \in [0, 3])$
in which a photon exists in the $k$th time slot.
When pure state $\ket{k}$ is launched into the 2-bit delay MZI,
the output state at port $p_{2x}$ is $\op{M}_{2x} \ket{k}$,
where generalized measurement operator $\op{M}_{2x}$ is given by
\begin{equation}
\op{M}_{2x} = \frac{1}{2} \sum_{k=0}^3 \left( \ket{k} + e^{i \theta_2} \ket{k+2} \right) \bra{k}		.
\end{equation}
Similarly,
we can obtain the operators representing the measurements of each MZI at ports $p_{2y}, p_{1x}$, and $p_{1y}$ as follows.
\begin{eqnarray}
\op{M}_{2y} &=& \frac{1}{2} \sum_{k=0}^3 \left( - \ket{k} + e^{i \theta_2} \ket{k+2} \right) \bra{k}		,	\\
\op{M}_{1x} &=& \frac{1}{2} \sum_{k=0}^5 \left( \ket{k} + e^{i \theta_1} \ket{k+1} \right) \bra{k}		,	\\
\op{M}_{1y} &=& \frac{1}{2} \sum_{k=0}^5 \left( -\ket{k} + e^{i \theta_1} \ket{k+1} \right) \bra{k}		.
\end{eqnarray}

Photon detectors D1 and D2 detect a photon at different detection times, $t_l$, for $l \in [0, 6]$,
which correspond to the projection measurements $\op{M}_D = \ket{l} \! \bra{l}$.
Therefore,
the expected value $n^E_{D1 l \theta_1 \theta_2}$ of the photons detected by D1 at time $t_l$ is given by
\begin{eqnarray}
n^E_{D1 l \theta_1 \theta_2} &=& N \mathrm{Tr} \left(
	\op{M}_D
		\op{M}_{1x}
			\op{M}_{2x}
				\op{\rho}
			\op{M}_{2x}^{\dag}
		\op{M}_{1x}^{\dag}
	\op{M}_D^{\dag}
\right)		\\
	&=& N \mathrm{Tr} \left( \op{E}_{l \theta_1 \theta_2}^{D1} \op{\rho} \right)		,	\label{eq:E(Count)ltt}
\end{eqnarray}
where we define the element of the positive operator valued measure
$\op{E}_{l \theta_1 \theta_2}^{D1} = \op{M}_{2x}^{\dag} \op{M}_{1x}^{\dag} \op{M}_D^{\dag} \op{M}_D \op{M}_{1x} \op{M}_{2x}$.
The element of the positive operator valued measure for D2 is similarly defined as 
$\op{E}_{l \theta_1 \theta_2}^{D2} = \op{M}_{2y}^{\dag} \op{M}_D^{\dag} \op{M}_D \op{M}_{2y}$.
To see what the measurement is performed by $\op{E}_{l \theta_1 \theta_2}^{DX}$ for $DX \in \{D1, D2\}$,
it is convenient to estimate the simplified forms of $\op{M}_D \op{M}_{1x} \op{M}_{2x}$ and $\op{M}_D \op{M}_{2y}$.
Fortunately,
$\op{M}_D$ is the projection onto the $l$th time slot for output states;
thus, they have the simplified forms as $w_{DX l} \ket{l} \bra{\psi^{DX}_{l \theta_1 \theta_2}}$,
where $w_{DX l}$ is a complex weight
and $\ket{\psi^{DX}_{l \theta_1 \theta_2}}$ is a normalized state in four-dimensional state.
All the simplified forms of the measurement operators are summarized in \tref{tab:OpList}.
Therefore,
$\op{E}_{l \theta_1 \theta_2}^{DX}$ returns the measurement result by the projector $\ket{\psi^{DX}_{l \theta_1 \theta_2}} \bra{\psi^{DX}_{l \theta_1 \theta_2}}$,
excluding the difference in weight $|w_{DX l}|^2$.
The simplified forms are easier to understand,
but the multiplication forms like $\op{M}_D \op{M}_{1x} \op{M}_{2x}$ are more convenient for expanding the dimension
or compensating for the imperfections due to measurement equipment as described later.

\begin{table}
	\caption{\label{tab:OpList}Measurement operators at different detection times and detector.}
	\begin{indented}
	\lineup
	\item[]
		\begin{tabular}{@{}ccr@{}r@{}r@{}r@{}r@{}r}
			\br
			Detector & Detection time & \multicolumn{6}{c}{Measurement operator}	\\
			\mr
			D1 & $t_0$ & $\frac{1}{4} \ket{0}$ & & $\bra{0}$ &&& \\
			 & $t_1$ & $\frac{1}{4} \ket{1}$ & $($ & $\bra{1}$ & $+e^{i \theta_1}\bra{0}$ && $)$	\\
			 & $t_2$ & $\frac{1}{4} \ket{2}$ & $($ & $\bra{2}$ & $+e^{i \theta_1}\bra{1}$ & $+e^{i \theta_2}\bra{0}$ & $)$	\\
			 & $t_3$ & $\frac{1}{4} \ket{3}$ & $($ & $\bra{3}$ & $+e^{i \theta_1}\bra{2}$ & $+e^{i \theta_2}\bra{1}$ & $+e^{i (\theta_1 + \theta_2)}\bra{0})$	\\
			 & $t_4$ & $\frac{1}{4} \ket{4}$ & $($ && $e^{i \theta_1}\bra{3}$ & $+e^{i \theta_2}\bra{2}$ & $+e^{i (\theta_1 + \theta_2)}\bra{1})$	\\
			 & $t_5$ & $\frac{1}{4} \ket{5}$ & $($ &&& $e^{i \theta_2}\bra{3}$ & $+e^{i (\theta_1 + \theta_2)}\bra{2})$	\\
			 & $t_6$ & $\frac{1}{4} \ket{6}$ & $($ &&&& $e^{i (\theta_1 + \theta_2)}\bra{3})$	\\
			\cline{2-8}
			D2 & $t_0$ & $-\frac{1}{2} \ket{0}$ && $\bra{0}$ &&&		\\
			 & $t_1$ & $-\frac{1}{2} \ket{1}$ && $\bra{1}$ &&&		\\
			 & $t_2$ & $-\frac{1}{2} \ket{2}$ & $($ & $\bra{2}$ && $-e^{i \theta_2}\bra{0}$ & $)$	\\
			 & $t_3$ & $-\frac{1}{2} \ket{3}$ & $($ & $\bra{3}$ && $-e^{i \theta_2}\bra{1}$ & $)$	\\
			 & $t_4$ & $-\frac{1}{2} \ket{4}$ & $($ &&& $-e^{i \theta_2}\bra{2}$ & $)$	\\
			 & $t_5$ & $-\frac{1}{2} \ket{5}$ & $($ &&& $-e^{i \theta_2}\bra{3}$ & $)$	\\
			\br
		\end{tabular}
	\end{indented}
\end{table}

As in the QST for qubits,
we need to rotate $\theta_1$ and $\theta_2$ to complete the QST for qudits.
We use the same combinations of phase differences $\theta_1$ and $\theta_2$ utilized
for the time-energy entangled qudits \cite{Richart2014}.
The total Hilbert space of the time-energy entangled qudits is spanned by two different logical qubits.
One is the qubit defined by the short and the long arms of the 1-bit delay MZI,
and the other is the qubit defined by the short and the long arms of the 2-bit delay MZI.
Therefore,
the high-dimensional QST is performed by the combination of the QST for logical qubits.
Setting the phase differences between the arms at $0$ and $\pi/2$ corresponds
to the measurements by the Pauli matrices $\sigma_x$ and $\sigma_y$ \cite{Nielsen2010} for logical qubits, respectively.
Therefore, combinations of phase differences $(\theta_1, \theta_2) = (0,0), (0, \pi/2), (\pi/2, 0)$, and $(\pi/2, \pi/2)$
are sufficient to obtain the information about the phase of the qudits.

On the other hand,
QST for qubits usually requires a measurement corresponding to the Pauli matrix $\sigma_z$,
which implies that it requires measurements without interference.
The measurement corresponding to $\sigma_z$ for both the logical qubits are performed by D2 at $t_0, t_1, t_4$ and $t_5$,
because the states $\ket{\psi^{D2}_{l \theta_1 \theta_2}}$ at these times are single time-bin basis states that correspond to eigenstates of $\sigma_z$.
However,
we need to prepare not only a $\sigma_z \otimes \sigma_z$ measurement for logical qubits that doesn't completely interfere
but also measurements that partially interfere like a $\sigma_z \otimes \sigma_x$ measurement.
From this point,
the measurements by D1 at different detection times play an important role in the proposed scheme,
because the interference pattern of the measurement $\op{E}_{l \theta_1 \theta_2}^{D1}$ depends on detection time $t_l$ as shown in \fref{fig:CMZI} and \tref{tab:OpList}.
In other words,
the combination of the time-bin basis constituting $\ket{\psi^{D1}_{l \theta_1 \theta_2}}$ varies depending on the detection time.
The measurement at $t_0$ by D1 corresponds to the projection onto the single time-bin basis $\ket{0}$,
the measurement at $t_1$ by D1 corresponds to the projection onto a superposition of $\ket{0}$ and $\ket{1}$, and so on.

Considering these characteristics of $\op{E}_{l \theta_1 \theta_2}^{DX}$ described above,
it is expected that the QST for time-bin qudits can be performed only by switching $\theta_1$ and $\theta_2$,
which is confirmed by comparing \eref{eq:ConceptOfQST} and \eref{eq:E(Count)ltt}
and by estimating the rank of $A_{ij}$.
The proposed scheme can be extended to general $d$-dimensional QST by adding extra MZIs.
The number of the MZIs for $d$-dimensional QST is $K$ given by $\lceil \log_2 d \rceil$,
where $\lceil x \rceil$ is the ceiling function for $x \in \mathbb{R}$.
The $K$ delay MZIs have different delay times $2^{i-1} T$ and phase differences $\theta_i$ for $1 \leq i \leq K$.
Each $\theta_i$ takes $0$ and $\pi / 2$ independently;
thus, the number of measurement settings scales linearly with $d$.

It should be noted that we can implement QST for time-bin qudits without D2,
which is confirmed from the rank of $A_{ij}$.
However,
D2 not only detects the photon which would be lost without it
but also collects information different from that obtained by D1.
For example,
D1 cannot implement the measurement corresponding to the projection onto $\ket{1}$, which D2 can.
This implies that D2 observes the same state from a different angle on the high-dimensional Bloch sphere.
Therefore, the addition of D2 effectively improves the accuracy of the QST in the same measurement time.

\subsection{\label{sec:LossComp}Compensation for imperfections}

The measurements described in subsection \ref{sec:BasicQST} are ideal ones without imperfection.
In practice,
there are no ideal 50 : 50 beam splitters and no photon detectors with $100 \%$ detection efficiency.
Furthermore, when we utilize delay MZIs made with planar light wave circuit technology (PLC),
the difference in the optical path length between the long and the short arms causes imperfection due to medium loss.
However,
the following modifications of the measurement operators can compensate for such imperfections:
\begin{eqnarray}
\op{M}_{2x} &=& \frac{\sum_{k=0}^3 \left( \ket{k} + \sqrt{\mathit{\Delta} \eta_{2x} } e^{i \theta_2} \ket{k+2} \right) \bra{k}}{\sqrt{2\left( 1+\mathit{\Delta} \eta_{2x} \right)}} 		,	\label{eq:CompM2x}	\\
\op{M}_{2y} &=& \frac{\sum_{k=0}^3 \left( - \ket{k} + \sqrt{\mathit{\Delta} \eta_{2y} } e^{i \theta_2} \ket{k+2} \right) \bra{k}}{\sqrt{2\left( 1+\mathit{\Delta} \eta_{2y} \right)}} 		,	\label{eq:CompM2y}	\\
\op{M}_{1x} &=& \frac{\sum_{k=0}^5 \left( \ket{k} + \sqrt{\mathit{\Delta} \eta_{1x} } e^{i \theta_1} \ket{k+1} \right) \bra{k}}{\sqrt{2\left( 1+\mathit{\Delta} \eta_{1x} \right)}} 		,	\label{eq:CompM1x}	\\
\op{M}_{1y} &=& \frac{\sum_{k=0}^5 \left( -\ket{k} + \sqrt{\mathit{\Delta} \eta_{1y} } e^{i \theta_1} \ket{k+1} \right) \bra{k}}{\sqrt{2\left( 1+\mathit{\Delta} \eta_{1y} \right)}} 		,	\label{eq:CompM1y}	\\
\op{E}_{l \theta_1 \theta_2}^{D1} &=& \mathit{\Delta} \eta_{1} \op{M}_{2x}^{\dag} \op{M}_{1x}^{\dag} \op{M}_D^{\dag} \op{M}_D \op{M}_{1x} \op{M}_{2x}	,	\label{eq:CompED1}	\\
\op{E}_{l \theta_1 \theta_2}^{D2} &=& \op{M}_{2y}^{\dag} \op{M}_D^{\dag} \op{M}_D \op{M}_{2y}	,	\label{eq:CompED2}
\end{eqnarray}
where $\mathit{\Delta} \eta_{2x}, \mathit{\Delta} \eta_{2y}, \mathit{\Delta} \eta_{1x}, \mathit{\Delta} \eta_{1y},$ and $\mathit{\Delta} \eta_{1}$ are relative transmittances.
Relative transmittances are the ratios between the transmittances depending on the optical paths and detectors.
We utilizes the relative values rather than absolute ones for experimental and theoretical convenience.
The use of the relative values decreases the expected value of the total photon number $N$ obtained by QST;
thus, it is not an accurate modification in this sense.
However,
the expected density operator $\op{\rho}$ will not change because $\op{\rho}$ is determined by the relative values of the photon counts.
Therefore,
the use of the relative values is justified for the purpose of QST.

\subsection{\label{sec:MLE}Maximum likelihood estimation}

As we mentioned in subsection \ref{sec:BasicQST},
QST for time-bin qudits can be performed by linear conversion of \eref{eq:ConceptOfQST}.
However,
it is well known that
the density operator obtained by linear conversion does not often satisfy positivity,
which implies the estimated density operator is unphysical \cite{James2001}.
Maximum likelihood estimation (MLE) is often used to avoid this problem \cite{Agnew2011,James2001,Richart2014,Takesue2009}.
First,
we use another representation of $\op{\rho}$ to enforce positivity as follows:
\begin{eqnarray}
\op{\rho} &=& \frac{\op{R}^\dag \op{R}}{ \mathrm{Tr} \left( \op{R}^\dag \op{R} \right)}		,	\\
N &=& \mathrm{Tr} \left( \op{R}^\dag \op{R} \right)	,
\end{eqnarray}
where $\op{R}$ is an operator having a triangular form \cite{James2001}.
MLE is performed by finding $\op{R}$ that minimizes the likelihood function $L\left(\op{R}\right)$ given by
\begin{equation}
L\left(\op{R}\right) = \sum_j \left[ \frac{\left(n_j^M - n_j^E \right)^2}{n_j^E} + \ln n_j^E	\right]	,	\label{eq:MLE}
\end{equation}
where $n_j^M$ is the measured photon count and $n_j^E$ is the expected photon count in \eref{eq:ConceptOfQST}.
The summation over $j$ is calculated for $j$ indicating different measurements.
Note that we add $\ln n_j^E$ to the likelihood function given in \cite{James2001}.
The likelihood function is derived from the probability of obtaining a set of photon counts $n_j^M$,
which is given by
\begin{equation}
P = \frac{1}{N_{norm}} \prod_j \exp \left[ - \frac{\left(n_j^M - n_j^E \right)^2}{2 \sigma_j^2} \right]	,
\end{equation}
where $N_{norm}$ is the normalization constant and $\sigma_j \approx \sqrt{n_j^E}$ is the standard deviation for the $j$th measurement.
However,
the normalization constant $N_{norm}$ can be approximated by $\prod_j \sqrt{2\pi}\sigma_j$ with Gaussian approximation,
which leads to the additional term  $\ln n_j^E$.
To perform MLE according to \eref{eq:MLE},
we need to precisely map $n^E_{D1 l \theta_1 \theta_2}$ and $n^E_{D2 l \theta_1 \theta_2}$ to $n_j^E$
because the intrinsically same measurements exist in the measurement settings.
For example,
the measurement at $t_0$ by D1 corresponding to the projection onto $\ket{0}$ does not depend on $\theta_1$ and $\theta_2$.
For this purpose,
we introduce space $V_j$,
which satisfies the following conditions:
\begin{eqnarray}
^\forall \left(DX, l, \theta_1, \theta_2 \right) \in V_j \ , \ ^\forall \left( DX', l', \theta_1', \theta_2' \right) \in V_{j'}	\nonumber\\
\frac{\op{E}_{l \theta_1 \theta_2}^{DX}}{\mathrm{Tr} \left( \op{E}_{l \theta_1 \theta_2}^{DX} \right)}
=
\frac{\op{E}_{l' \theta_1' \theta_2'}^{DX'}}{\mathrm{Tr} \left( \op{E}_{l' \theta_1' \theta_2'}^{DX'} \right)}
\qquad
\mbox{for}
\qquad
j=j'	,		\label{eq:Vcondition1}
\\
\frac{\op{E}_{l \theta_1 \theta_2}^{DX}}{\mathrm{Tr} \left( \op{E}_{l \theta_1 \theta_2}^{DX} \right)}
\neq
\frac{\op{E}_{l' \theta_1' \theta_2'}^{DX'}}{\mathrm{Tr} \left( \op{E}_{l' \theta_1' \theta_2'}^{DX'} \right)}
\qquad
\mbox{for}
\qquad
 j \neq j'	.	\label{eq:Vcondition2}
\end{eqnarray}
Space $V_j$ is numerically generated via a comparison according to \eref{eq:Vcondition1} and \eref{eq:Vcondition2}.
By utilizing $V_j$,
we can map $n^E_{D1 l \theta_1 \theta_2}$ and $n^E_{D2 l \theta_1 \theta_2}$ to $n_j^E$ as follows:
\begin{eqnarray}
n_j^E &=& \sum_{\left(DX, l, \theta_1, \theta_2 \right) \in V_j} n^E_{D1 l \theta_1 \theta_2}		\\
&=& N \mathrm{Tr} \left( \op{E}_{j} \op{\rho} \right)	,
\end{eqnarray}
where $\op{E}_{j} = \sum_{\left(DX, l, \theta_1, \theta_2 \right) \in V_j} \op{E}_{l \theta_1 \theta_2}^{DX}$.
Similarly, we obtain $n_j^M$,
and now we can perform the QST for time-bin qudits by MLE.

\subsection{\label{SumOfProc}Summary}

Here,
we summarize the proposed QST procedure.

First,
we measure the relative transmittances
$\mathit{\Delta} \eta_{2x}, \mathit{\Delta} \eta_{2y}, \mathit{\Delta} \eta_{1x}, \mathit{\Delta} \eta_{1y},$ and $\mathit{\Delta} \eta_{1}$,
with which we estimate the measurement operators $\op{E}_{l \theta_1 \theta_2}^{DX}$ according to
\eref{eq:CompM2x}--\eref{eq:CompED2}.
Then,
we generate space $V_j$ from $\op{E}_{l \theta_1 \theta_2}^{DX}$ according to \eref{eq:Vcondition1} and \eref{eq:Vcondition2}
and prepare $\op{E}_{j} = \sum_{\left(DX, l, \theta_1, \theta_2 \right) \in V_j} \op{E}_{l \theta_1 \theta_2}^{DX}$.

Next,
we perform photon count measurement
by switching combinations of phase differences $(\theta_1, \theta_2) = (0,0), (0, \pi/2), (\pi/2, 0)$, and $(\pi/2, \pi/2)$
and obtain $n^M_{DX l \theta_1 \theta_2}$.
After the measurement,
$n^M_{DX l \theta_1 \theta_2}$ is reduced into $n_j^M$ by using space $V_j$.

Finally,
we find $\op{R}$ minimizing the likelihood function $L\left(\op{R}\right)$ with $n_j^M$ and $\op{E}_{j}$
and obtain the reconstructed density operator $\op{\rho}$.
When we perform the QST for the multi-photon state,
we extend the procedure as in \cite{James2001,Thew2002}
by replacing $\op{E}_{l \theta_1 \theta_2}^{DX}$ and $n^M_{DX l \theta_1 \theta_2}$ with its tensor production and coincidence count,
respectively.

\section{Experimental setup}

\begin{figure}[htbp]
\centering
\includegraphics[width=\linewidth]{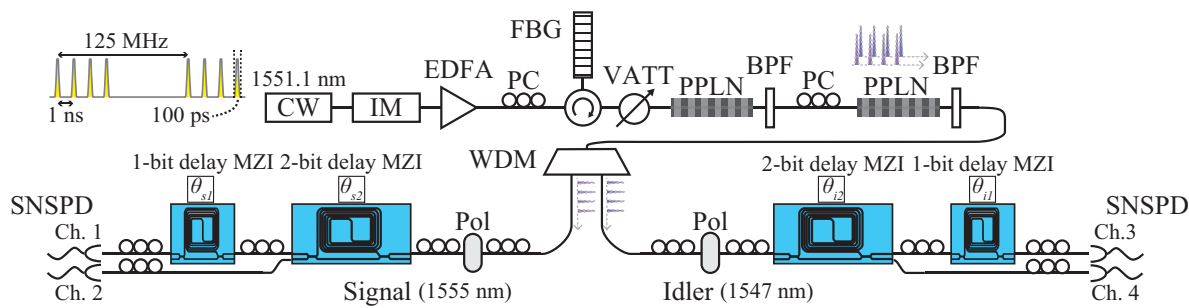}
\caption{Experimental setup.
CW: Continuous wave laser.
IM: Intensity modulator.
EDFA: Erbium-doped fiber amplifier.
PC: Polarization controller.
FBG: Fiber Bragg grating filter.
VATT: Optical variable attenuator.
PPLN: Periodically poled lithium niobate waveguide.
BPF: Optical band-pass filter.
WDM: Wavelength demultiplexing filter.
Pol: Polarizer.
2-bit delay MZI, 1-bit delay MZI (Delay Mach-Zehnder interferometers were fabricated using PLC technology.)
SNSPD: Superconducting nanowire single-photon detector.
}
\label{fig:ExpSetup}
\end{figure}

\Fref{fig:ExpSetup} shows the experimental setup.
First,
we generate a continuous-wave light with a wavelength of 1551.1 nm and a coherence time of $\sim$10 $\mu$s,
which is modulated into four-sequential pulses by an intensity modulator.
The repetition frequency, the temporal interval, and the pulse duration are 125 MHz, 1 ns, and 100 ps, respectively.
These pulses are amplified by an erbium-doped fiber amplifier (EDFA),
and then the average power of the pulses are adjusted by an optical variable attenuator.
They are launched into a periodically poled lithium niobate (PPLN) waveguide,
where 780-nm pump pulses are generated via second harmonic generation.
The 780-nm pump pulses are launched into another PPLN waveguide to generate a four-dimensional maximally entangled state through spontaneous parametric down-conversion.
A fiber Bragg grating filter and two optical band-pass filters are located after the EDFA and the PPLN waveguides, respectively.
The fiber Bragg grating filter eliminates amplified spontaneous emission noise from the EDFA,
and the first and the second band-pass filters eliminate the 1551.1- and the 780-nm pump pulses, respectively.
The generated entangled photons are separated by a wavelength demultiplexing filter into a signal and an idler photon whose wavelengths are 1555 and 1547 nm, respectively.
Each separated photon is launched into the cascaded MZIs followed by two superconducting nanowire single-photon detectors (SNSPDs),
where the QST described in \sref{sec:QSTDetail} is performed.
The cascaded MZIs are composed of a 2-bit delay MZI and a 1-bit delay MZI fabricated by using PLC technology.
The phase differences of the 2- and 1-bit delay MZIs are controlled via the thermo-optic effect caused by electrical heaters attached to the waveguides.
Each MZI shows a $>20$-dB extinction ratio thanks to the stability of the PLC \cite{Takesue2005, Honjo2004}.
Polarization controllers and polarizers are located in front of each MZI to operate the MZIs for one polarization.
Channels 1 and 2 (3 and 4) of the SNSPDs are connected to the 1- and the 2-bit delay MZIs for the signal (idler) photon, respectively.
The photon detection events from the SNSPDs are recorded by a time-interval analyzer
and analyzed by a conventional computer.
The detection efficiencies of the SNSPDs for channels 1, 2, 3, and 4 are $40, 56, 34$, and $43$ \%, respectively,
and the dark count rate for all channels is $<30$ cps.

\section{Results}

\subsection{Measurement of relative transmittance}
\begin{figure}[htbp]
\centering
\includegraphics[width=\linewidth]{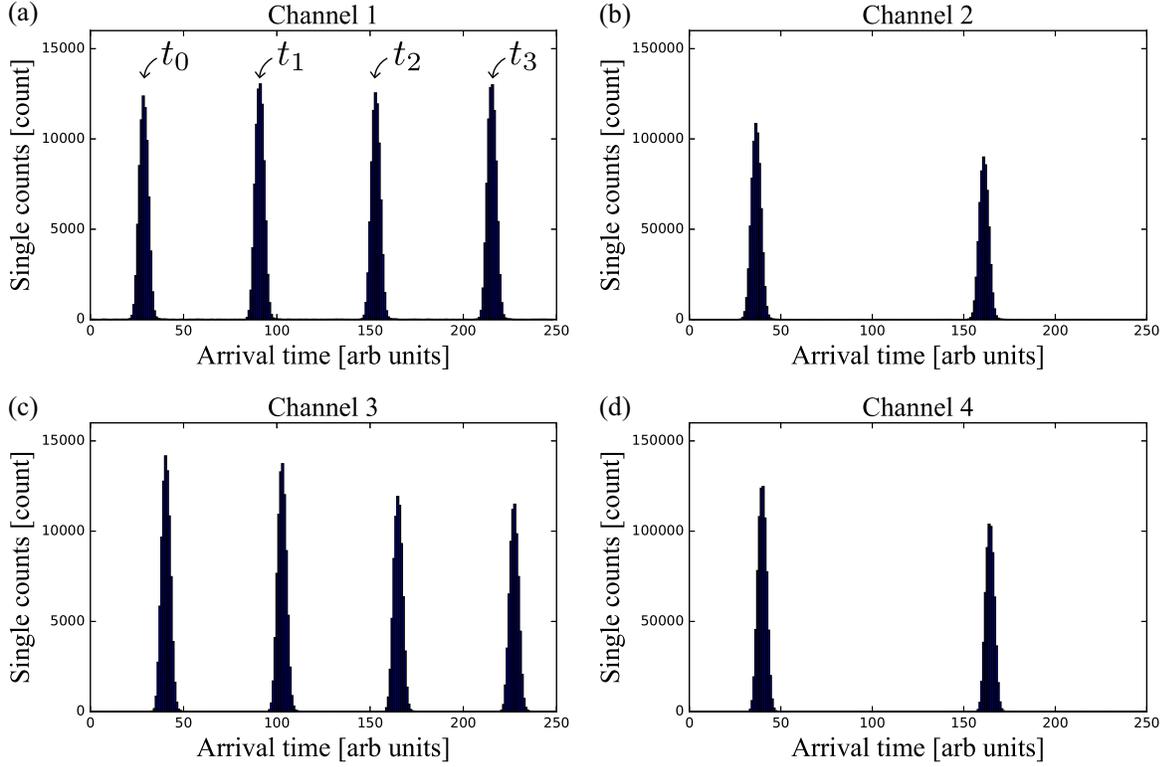}
\caption{Histograms of single counts for single photon generated by single pump pulse
for the detector's (a) channel 1, (b) channel 2, (c) channel 3, and (d) channel 4.}
\label{fig:DiffEta}
\end{figure}

We first measured the relative transmittances between the arms of the MZIs---$\mathit{\Delta} \eta_{2x}, \mathit{\Delta} \eta_{2y},$ and $\mathit{\Delta} \eta_{1x}$---for the signal and the idler photon.
To measure these values,
we generated a single pulse using the intensity modulator instead of four-sequential ones,
because the photons generated by the single pulse don't interfere at the MZIs.
\Fref{fig:DiffEta} shows the histograms of single photon counts for each detector channel.
The four peaks in \fref{fig:DiffEta}(a) and (c) correspond to the single counts for finding a photon in detection times $t_0$, $t_1$, $t_2$, and $t_3$, respectively.
Similarly,
the two peaks in \fref{fig:DiffEta}(b) and (d) correspond to the single counts for finding a photon in detection times $t_0$ and $t_2$, respectively.
We calculated the relative transmittances from these single counts.
For example,
single count $S^1_l$ at detection time $t_l$ for channel 1 satisfies the following relation:
\begin{equation}
S^1_0 : S^1_1 : S^1_2 : S^1_3 = 1 : \mathit{\Delta} \eta_{1x}^s : \mathit{\Delta} \eta_{2x}^s : \mathit{\Delta} \eta_{1x}^s \mathit{\Delta} \eta_{2x}^s	,
\end{equation}
where $\mathit{\Delta} \eta_{2x}^s$ and $\mathit{\Delta} \eta_{1x}^s$ are the relative transmittances for the signal photon.
Therefore,
the relative transmittances were estimated as
$\mathit{\Delta} \eta_{2x}^s = \left( S^1_2 + S^1_3 \right) / \left( S^1_0 + S^1_1 \right)$
and
$\mathit{\Delta} \eta_{1x}^s = \left( S^1_1 + S^1_3 \right) / \left( S^1_0 + S^1_2 \right)$.
Similarly,
we calculated the other relative transmittances,
which are summarized in \tref{tab:DiffEta}.
We didn't measure $\mathit{\Delta} \eta_{1y}$
because output port $p_{1y}$ was terminated
and thus didn't affect the result of our experiment.
The values summarized in \tref{tab:DiffEta} were utilized for the QST described in the next section.

\begin{table}
	\caption{\label{tab:DiffEta}Summary of the relative transmittance.}
	\begin{indented}
	\lineup
	\item[]
		\begin{tabular}{@{}crr}
			\br
			 & Signal & Idler	\\
			\mr
			$\mathit{\Delta} \eta_{2x}$ & 1.009\0 & 0.8495	\\
			$\mathit{\Delta} \eta_{2y}$ & 0.8300 & 0.8302	\\
			$\mathit{\Delta} \eta_{1x}$ & 1.063\0 & 0.9669	\\
			\br
		\end{tabular}
	\end{indented}
\end{table}

\subsection{QST for the time-bin entangled qudits}

\begin{figure}[htbp]
\centering
\includegraphics[width=\linewidth]{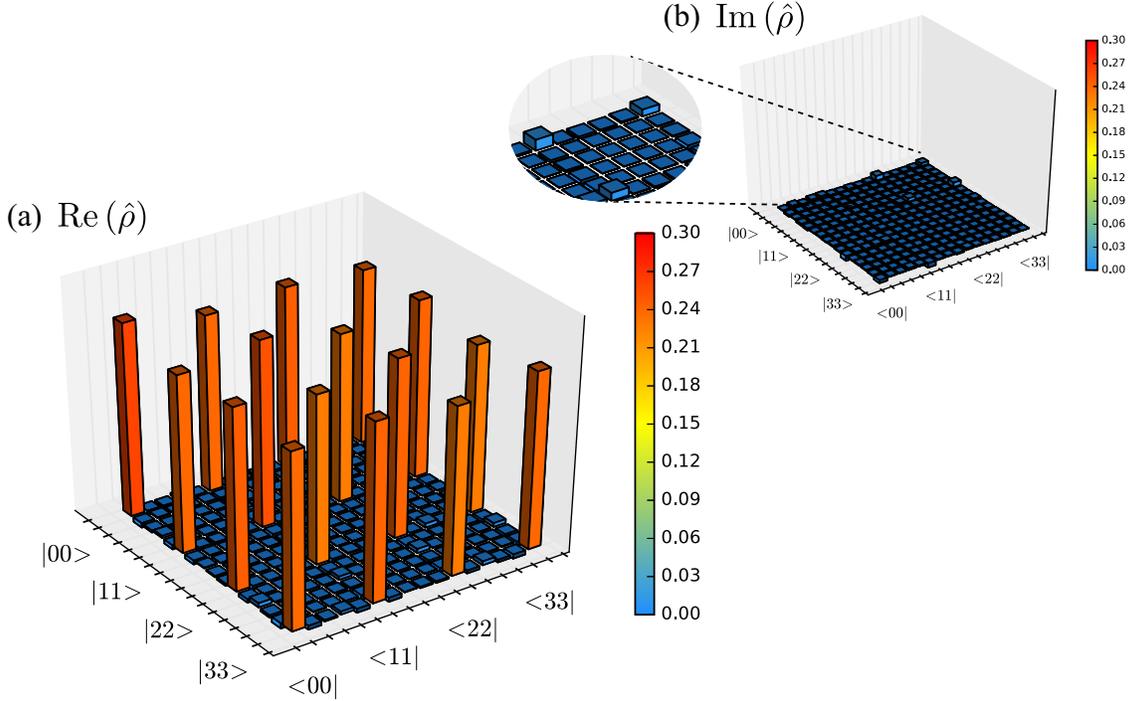}
\caption{(a) Real parts and (b) imaginary parts of measured density operator $\op{\rho}$.}
\label{fig:Rho}
\end{figure}

We then generated the four-dimensional maximally entangled state $\ket{\Psi_{MES}^4 (\phi)}$ by utilizing the four-sequential pump pulses.
The state is given by
\begin{equation}
\ket{\Psi_{MES}^4 (\phi)} = \frac{1}{2} \sum_{k=0}^3 \exp (i \phi k ) \ket{k}_s \otimes \ket{k}_i 	,
\end{equation}
where $\ket{k}_s$ and $\ket{k}_i$ denote the time-bin basis for the signal and idler photon, respectively,
and $\phi$ denotes the relative phase between the product states $\ket{k}_s \otimes \ket{k}_i$
due to the phases of the pump pulses for SPDC.
The pump pulses were generated from the CW laser;
thus, the phase is proportional to $k$ and determined by the frequency and the temporal interval of the time slots.
It should be noted that we can control the phase of the entangled state by modulating that of the pump pulses.
In our setup,
the CW laser had a coherence time of $\sim$10 $\mu$sec,
which implies that, in principle, we can extend the dimension of the entangled photons $d$ up to $10^3\sim10^4$.
The measured single photon count rates for detector channels 1, 2, 3, and 4 were
17.1, 72.4, 20.6, and 82.1 kcps, respectively.
From these single photon count rates,
the relative transmittances between the detectors $\mathit{\Delta} \eta_{1}$ for the signal and idler photon
were estimated to be 0.474 and 0.501, respectively.
The average photon number per qudit was 0.02,
and the measurement time for one measurement setting was 10 sec.
We employed coincidence counts for arbitrary combinations of detection times between the signal and the idler photon with 16 measurement settings,
with which the QST for a single qudit described in \sref{sec:QSTDetail} was extended to the QST for two qudits.

We performed the QST for the entangled qudits fifteen times.
\Fref{fig:Rho} shows one of the measured density operators $\op{\rho}$.
All measured coincidence counts and reconstructed operators in the fifteen trials are provided in the supplementary material.
Note that we utilized $\op{U}\op{\rho}\op{U}^\dag$ instead of $\op{\rho}$ so that the visualized operator would be close to $\ket{\Psi_{MES}^4 (0)}$,
where the local unitary operator $\op{U}$ for the signal photon is given by $\sum_k \exp (-i \phi' k) \ket{k}_s\bra{k}_s$.
Both the real and the imaginary parts of the measured operator showed characteristics close to $\ket{\Psi_{MES}^4 (0)}$,
and the elements of the operator that were 0 for $\ket{\Psi_{MES}^4 (0)}$ were suppressed.

\begin{table}
	\caption{\label{tab:FigOfMerit}Average quantities derived from measured $\op{\rho}$ for the fifteen experimental trials.
	The critical values to violate the CGLMP inequality are also summarized.}
	\begin{indented}
	\lineup
	\item[]
		\begin{tabular}{@{}crr}
			\br
			 & Measured & Critical \\
			\mr
			Fidelity & $F(\op{\rho}, \op{\sigma}) = \m $ 0.950 $\pm$ 0.003 & $> 0.710$ \\
			Trace distance & $D(\op{\rho}, \op{\sigma}) = \m $ 0.068 $\pm$ 0.003 & $< 0.290$ \\
			Linear entropy & $H_{lin}(\op{\rho}) = \m $ 0.093 $\pm$ 0.006 & $< 0.490$ \\
			Von Neumann entropy & $H_{vn}(\op{\rho}) = \m $ 0.343 $\pm$ 0.016 & $< 2.002$ \\
			\multirow{2}{*}{Conditional entropy} & $H_{c}(\op{\rho}|s) = - $ 1.654 $\pm$ 0.016 & $< 0.002$ \\
			 & $H_{c}(\op{\rho}|i) = - $ 1.653 $\pm$ 0.016 & $< 0.002$ \\
			\br
		\end{tabular}
	\end{indented}
\end{table}

To evaluate the measured operators more quantitatively,
we derived five figures of merit from $\op{\rho}$:
fidelity $F(\op{\rho}, \op{\sigma})$,
trace distance $D(\op{\rho}, \op{\sigma})$,
linear entropy $H_{lin}(\op{\rho})$,
von Neumann entropy $H_{vn}(\op{\rho})$,
and conditional entropy $H_{c}(\op{\rho}|X)$ \cite{Nielsen2010,James2001}.
Here, we employed the following definitions:
\begin{eqnarray}
F(\op{\rho}, \op{\sigma}) &=& \left[ \Tr \sqrt{ \sqrt{\op{\sigma}} \op{\rho} \sqrt{\op{\sigma}}}\right]^2	,	\\
D(\op{\rho}, \op{\sigma}) &=& \frac{1}{2} \Tr \sqrt{\left(\op{\rho} - \op{\sigma}\right)^2}	,	\\
H_{lin}(\op{\rho}) &=& 1 - \Tr \left( \op{\rho}^2 \right)	,	\\
H_{vn}(\op{\rho}) &=& - \Tr \left( \op{\rho} \log_2 \op{\rho}\right)	,	\\
H_{c}(\op{\rho}|X) &=& H_{vn}(\op{\rho}) - H_{vn}(\op{\rho}_X)	,
\end{eqnarray}
where $\op{\sigma}$ is given by$\ket{\Psi_{MES}^4 (\phi)} \bra{\Psi_{MES}^4 (\phi)}$ with $\phi$,
which maximizes $F(\op{\rho}, \op{\sigma})$ or minimizes $D(\op{\rho}, \op{\sigma})$,
$X \in \{s, i \}$ denotes the signal and idler photon, respectively,
and $\op{\rho}_X$ is the reduced density operator for $X$.
The average values of these quantities are summarized in \tref{tab:FigOfMerit}.
The errors in \tref{tab:FigOfMerit} were estimated as standard deviations in the fifteen experimental trials.
Therefore,
they included the statistical characteristics of the coincidence counts and all the effects due to the experimental imperfections as well.
The measured fidelity and trace distance showed that the reconstructed operators were close to the target state $\ket{\Psi_{MES}^4 (\phi)}$.
Note that this is the first time fidelity $>0.90$ has been reported for entangled qudits \cite{Agnew2011,Bernhard2013,Nowierski2015,Richart2014}.
The measured linear entropy and von Neumann entropy were low,
which implies that the reconstructed operators were close to the pure state and that small disturbances occurred in the proposed QST scheme.
Furthermore,
the measured conditional entropies were negative,
which confirmed that the signal and the idler photons were entangled \cite{Horodecki1996a,Horodecki1996}.

To evaluate the quality of entangled qudits,
many previous experiments employed the Collins-Gisin-Linden-Massar-Popescu (CGLMP) inequality test,
which is a generalized Bell inequality for entangled qudits \cite{Collins2002,Dada2011a}.
If we assume symmetric noise,
depolarized entangled state $\op{\rho}_{mix}$ is given by
\begin{equation}
\op{\rho}_{mix} = p \ket{\Psi_{MES}^4 (0)} \bra{\Psi_{MES}^4 (0)} + (1 - p) \frac{\op{I}_{16}}{16}	,		\label{eq:MixedMES}
\end{equation}
where $p$ is a probability and $\op{I}_{16}$ is the identity operator in the 16-dimensional Hilbert space.
The condition $p > 0.69055$ is a criterion to violate the CGLMP inequality.
Therefore,
the quantities derived from $\op{\rho}_{mix}$ with $p = 0.69055$ can be considered as the critical values for the evaluation of the entangled qudits.
These critical values are also summarized in \tref{tab:FigOfMerit},
which shows that all of the measured values satisfied the conditions to violate the CGLMP inequality.
Thus,
we confirmed that the proposed QST scheme based on cascaded MZIs successfully reconstructed the quantum density operator of the time-bin entangled qudits
with only 16 measurement settings.

\section{Conclusion}
We proposed QST for time-bin qudits based on cascaded MZIs,
with which the number of measurement settings scales linearly with dimension $d$.
We generated a four-dimensional maximally entangled time-bin state
and confirmed that the proposed scheme successfully reconstructed the density operator with only 16 measurement settings.
All the quantities derived from the reconstructed state were close to the ideal ones,
and the fidelity of 0.950 is the first time fidelity $>0.90$ has been achieved for entangled qudits.
We hope that our result will lead to advanced quantum information processing utilizing high-dimensional quantum systems.

\ack
We thank T. Inagaki and F. Morikoshi for fruitful discussions.

\section*{References}

\bibliographystyle{iopart-num}
\bibliography{sample}

\end{document}